# Noise Probe of the Dynamic Phase Separation in $La_{2/3}Ca_{1/3}MnO_3$

B. Raquet[†], A. Anane[*], S. Wirth, P. Xiong and S. von Molnár,

MARTECH, Florida State University, Tallahassee, Florida 32306-4351

**ABSTRACT**

Giant Random Telegraph Noise (RTN) in the resistance fluctuation of a macroscopic film of perovskite-type manganese oxide $La_{2/3}Ca_{1/3}MnO_3$ has been observed at various temperatures ranging from 4K to 170K, well below the Curie temperature ($T_C \approx 210K$). The amplitudes of the two-level-fluctuations vary from 0.01% to 0.2%. We discuss the origin of the RTN to be a dynamic mixed-phase percolative conduction process, where manganese clusters switch back and forth between two phases that differ in their conductivity and magnetization.

[*]E-mail : anane@martech.fsu.edu

The current interest in the mixed valence perovskite manganites was originally fueled by the rediscovery of colossal magnetoresistance (CMR) in certain members of this group of materials [1]. The study of the physics of the manganites has, however, progressed far beyond CMR alone. Manganites have offered a fertile ground for the study of spin-charge interactions and transport-magnetism correlations. It is unusual that the same solid state system can exhibit so many ground states, ranging from ferromagnetic-metallicity to charge-ordering and orbital-ordering [2,3]. This richness is due to competition between a variety of interactions of comparable strengths [4]. The energy balance can be so delicate that many have suggested that the ground state would not be homogeneous [5-7]. Indeed, a variety of experimental studies of the crystal and magnetic structures support the existence of both lattice [8] and magnetic polaronic [9,10] states near the ferromagnetic ordering temperature $T_C$. Furthermore, a multiphase description, where the insulator metal transition occurs via percolation has been put forward [11]. This picture of the coexistence of metallic and insulating phases, even in samples optimally-doped to have the highest $T_C$, has recently been confirmed experimentally by Scanning Tunneling Microscopy (STM) [12, 13].

Random Telegraph Noise (RTN) results from transitions between two states of a switching entity called “fluctuator”. The two states could be electrical or magnetic in origin. Each state has a different conductivity, thus makes a different contribution to the overall conductivity; hence the switching results in discrete jumps in the overall sample resistance. Studying the temperature and field dependence of the RTN (switching behavior) provides insight into the electrical and magnetic nature of the fluctuators. Therefore, RTN is an effective tool for probing the dynamic behavior of an electrically inhomogeneous system, such as the manganites.

In this letter, we report on the first observation of giant RTN in the resistance of a $La_{2/3}Ca_{1/3}MnO_3$ film. We have examined in detail the RTN as a function of temperature and magnetic field. We use a statistical analysis of the life-times of the Two Level Fluctuations (TLF) to gain insight into the microscopic electronic and magnetic state of this system. At low temperature (below 30K) the TLF is well described by a thermally activated two-level model. At higher temperature (between 60K and 170K) we observed critical effects of the temperature on the life-times of the TLF. Our results provide evidence for a dynamic phase separation in this system. More importantly, for the first time, we are able to give a quantitative estimate of the relative stability of these phases.

A 0.4μm thick film was grown by pulsed laser deposition on a $LaAlO_3$ (100) substrate. It was patterned into a long four-probe geometry of 200μm in length and 2 μm in width using a combination of optical lithography and wet chemical etching. The noise measurements were performed with a four probe DC technique as described in Ref. 14. Care was taken to ensure that spurious noise does not contribute to our results.

We have observed RTN at various temperatures below $T_C$. From 4K to 170K, the resistance fluctuations alternate between an anomalous non-Gaussian behavior to giant RTN as shown in Fig.1. The fractional resistance steps of the RTN vary from 0.01% to greater than 0.2%, which is surprisingly high for an almost macroscopic sample.

The TLF process can be modeled as an asymmetrical two-state system with an energy barrier separating the two states (Fig.2a). Obviously, here the switching is thermally activated as indicated in Fig. 2b. The average time $\tau_i$ spent in the $i$th state can then be described by the Arrhenius law:

$$\tau_i = \tau_{0,i} \exp\left(\frac{E_i}{k_B T}\right) \qquad (1)$$

where $E_i$ is the height of the barrier to escape from state $i$, $k_B$ is the Boltzmann constant and $\tau_{0,i}$ is a microscopic constant that depends on the details of the coupling between the thermal bath and the fluctuator.

We would like to emphasize that in the temperature range 4K to 180K the noise has a strong non-Gaussian character, *i.e.*, the resistance fluctuations are dominated by "few" fluctuators. The features observed in the traces in this temperature range are reminiscent of a superposition of "few" TLF processes with different characteristic life-times (traces recorded at 38K and 156K, Fig. 1). In our experiment, the accessible time-window ranges from 10 milliseconds to a few minutes. If the average life-time of a fluctuator (Eq. 1) falls outside of the experimental window, then it is not directly observable. The observable fluctuators are thus selected by setting the working temperature. In fact, at certain temperatures only a single fluctuator dominates, giving rise to the RTN (Fig.1). For some of these TLF processes we were able to study the temperature and magnetic field dependence of the average life-time ($\tau_i$). It is this study that constitutes the focus of this letter.

In Figure 2.b, the temperature dependence of the mean life-times for a TLF system active between 10K and 20K is presented. Although very limited in temperature range, a fit to Eq.(1) for the average life-time of the low-resistance state $\tau_{down}$ (the high-resistance state $\tau_{up}$) yields a measure of the energy barrier height, $E_{down}/k_B = 380K \pm 30K$ ($E_{up}/k_B = 270K \pm 20K$). The attempt-time is found to be $\tau_{0,down} = 3 \cdot 10^{-11}$s ($\tau_{0,up} = 2 \cdot 10^{-8}$s). Since the phonon spectrum is *a priory* the same for the two sates, the discrepancy between the attempt-times $\tau_{0,down}$ and $\tau_{0,up}$ is likely due to a slight temperature dependence of the energy barrier heights. This point will be discussed in more details below.

The magnetic field dependence of the noise reveals the magnetic nature of the switching entities (Fig. 2c). The effect of the field is to systematically stabilize the low-resistance state to the detriment of the high-resistance state. Within the framework of the TLF model, the effect of the magnetic field can be accounted for by including a field-dependent energy barrier in Eq.1, $E_i(\mathbf{H}) = E_i(0) + \Delta\mathbf{m}_i \cdot \mathbf{H}$, where $\Delta\mathbf{m}_i = \mathbf{m}_i - \mathbf{m}_v$, with $\mathbf{m}_i$ being the magnetic moment associated with the fluctuator in the state *i*, $\mathbf{m}_v$ being the magnetic moment of the virtual state (at the top of the energy barrier) and **H** being the applied field. From Figure 2c, we infer $\Delta m_{down} \approx -\Delta m_{up} \approx 550\mu_B$, where $\mu_B$ is the Bohr magneton.

Magnetic noise has been reported in various systems [15,16], and has most commonly been attributed to magnetic domain fluctuations. In our case, however, this scenario can be readily ruled out: *1)* we have observed RTN in fields higher than the demagnetization field (0.6T). Such fields tend to suppress any magnetic domains and therefore any magnetic domain fluctuations; *2)* if we assume the magnetic entity responsible for the resistance change to be a magnetic domain, its volume derived from the known $\Delta m_i$ values would be of the order of $(2nm)^3$, many orders of magnitudes too small to account for a relative variation of the resistance of 0.2% [17,18].

As the temperature rises above ~ 60K, we observed a striking change in the temperature dependence of the life-times of the TLF. As shown in Fig.3a, for a fluctuator active around 109K, we typically observe that over a narrow temperature range of ~ 1K, the life-times of the low-resistance and the high-resistance state change by more than an order of magnitude. Nevertheless, in contrast to the low temperature case, where both life-times decrease with increasing T, here the life-time of the high-resistance state actually increases with rising T. In fact, we observe a dramatic inversion of the occupation probability between the two states over this temperature range. A fraction of a degree K above the equiprobability temperature, the occupation probability of the low-resistance state becomes negligible compared to that for the now much more stable high-resistance state. In the analysis of the low temperature data in Eq.(1), the heights of the energy barriers are assumed to be temperature independent, and the temperature dependence of $\tau_i$ results purely from thermal activation. Namely there is no change in the energy configuration with varying T. This is clearly not the case any longer at high T, although we still expect the dynamics to be dominated by thermal activation and Eq.(1) remains valid for a given energy configuration (*i.e.* at a given temperature). We thus extend the TLF model by introducing a temperature dependent energy barrier $E_i(T,H)$. By inverting Eq.(1) we can estimate the temperature dependence of the energy barrier from the measured life time $\tau_i$ .

$$E_i(T,H) = k_B T \ln\left(\frac{\tau_i(T,H)}{\tau_{0,i}}\right) \tag{2}$$

Over the studied temperature range (~1K), we can neglect the temperature dependence of $\tau_{0,i}$. Moreover, the value of $\tau_{0,i}$ is mainly a logarithmic offset in Eq.2.

Figure 3b displays the temperature dependence of the activation energies deduced from Fig 3a. We assume that in this narrow temperature range, the temperature dependence of the free energy can be described by a functional, $F(\sigma)$, dependent on a dimensionless configuration parameter $\sigma$. In analogy to the Landau-Ginzburg model for first order phases transitions [19], we write $F(\sigma)$ as a polynomial expansion in $\sigma$. The simplest possible form, putting aside any symmetry argument, is a fourth order polynomial. The fit of the local extrema of $F(\sigma)$ reproduce well the temperature dependence for the energy barriers of both the low-and high-resistance states (Fig. 3 caption). Similar to the low temperature case, the effect of a magnetic field, at high temperature, is to stabilize the low-resistance state. The corresponding net magnetization difference is $\Delta m_{down} \approx -\Delta m_{up} \approx 2500\mu_B$.

Except when a single fluctuator dominates, the power spectral density of the resistance fluctuation has mainly a 1/*f* dependence, in good agreement with previous studies [17,20,21]. The normalized noise level shows a 4 orders of magnitude increase when the sample temperature is increased from 4K to 100K and then decreases sharply at higher temperature to the background level, whereas the resistivity maximum occurs at 210K, well above the noise maximum. It is worth mentioning that above 180K we have seen no evidence for the RTN, and the non-Gaussian character of the resistance fluctuations is much weaker.

Obviously the large noise level below 180K is due to the presence of two-level fluctuators with large $\Delta R/R$ values. We have established, however, that those fluctuators, even though they have a strong magnetic character, cannot originate from magnetic domains. Instead, we believe that the rapid variation of the energy configuration with temperature, as inferred from the RTN at high temperature (Fig.3), suggest the dynamic coexistence of two phases: a ferromagnetic metallic phase and a phase with relatively depressed magnetic and electrical properties. The RTN occurs when a cluster, the fluctuator, switches back and forth between the two phases. The magnetic field will always stabilize the ferromagnetic state, which is the low-resistance one. At low temperature the conductivity is dominated by the ferromagnetic metallic phase; by increasing the temperature, some clusters will switch to the depressed state, increasing the total resistance of the sample by the RTN quanta. We surmise, therefore, that the conduction is a mixed-phase percolation process, consistent with the phenomenological model of Jaime *et al* [22]. These authors developed a model of coexisting metallic and thermally activated (polaronic) transport. Conduction near $T_C$ is activated, whereas the metallic character dominates at low temperature (far from $T_C$). In this picture, the location of the noise level peak well below $T_C$ and its surprising amplitude is a direct consequence of the mixed-phase: near the percolation threshold for the metallic state, the conduction is dominated by the narrowest current paths. A few switching clusters located in these critical bonds will have dramatic effect on the overall connectivity of the metallic network, which results in a large increase of the noise level [23,24].

Finally, to address the issue of the intrinsic or extrinsic nature of the observed phase separation, we emphasize that the final state was never exactly the same upon thermal cycling: the RTN could disappear altogether and, if present, displays different characteristics, *i.e*, different $\tau_{0,i}$, $E_i$. This leads us to conclude that the mixed phase is not related to any chemical inhomogeneity or physical disorder but rather to a statistical probability to have the sample in a given state out of many possible configurations with comparable energies.

In conclusion, we have shown for the first time direct evidence of phase separation in $La_{2/3}Ca_{1/3}MnO_3$ by a transport measurement. The present study gives a rough estimate of ~100K for the energy difference between the two states at low temperature. A typical size of the switching clusters can be estimated from the net magnetization difference $\Delta m$. A simple model (*to be published*) yields an order of magnitude volume of $(20nm)^3$ at 20K which involves $10^5$ Mn atoms.

**Acknowledgements**

We thank L. P. Gor'kov for illuminating discussion. This work has been supported by DARPA and the office of Naval Research under Contract No. ONR-N00014-96-1-0767.

REFERENCES:

† Permanent address: Laboratoire de Physique de la Matière Condensée de Toulouse & Laboratoire national des champs magnétiques pulsés, INSA, Av. de Rangueil. 31077 Toulouse, France.

FIGURE CAPTIONS:

Fig. 1 : Resistance *vs* time at different temperatures.The noise alternates from a strongly non-Gaussian fluctuation-type to a Random Telegraph Noise (RTN) with $\Delta R/R$ ranging from 0.01% to 0.2%. The features observed in the traces recorded at 38K and 156K are reminiscent of a superposition of "few" Two Level Fluctuations (TLF) processes with different characteristic life-times. At certain temperatures only a single fluctuator dominates, giving rise to the RTN at 15K, 71K, 109K and 167K. There is no evidence for the RTN above 180K (the Curie temperature $T_C \approx 210K$).

Fig. 2 : (**a**) Schematic representation of an asymmetric double-well model used to describe the fluctuation process. (**b**) The measured average life-times for the high-resistance state ($\tau_{up}$) and the low-resistance state ($\tau_{down}$) are plotted versus the reciprocal temperature; (**c**) $\tau_{up}$ and $\tau_{down}$ versus magnetic field. In this temperature range, increasing the temperature decreases the life-time of both states while an applied magnetic field increasingly favors the low-resistance state to the detriment of the high-resistance one. The solid lines are fits to $\tau_i = \tau_{0,i} \exp\left(\frac{E_i(H=0)+\Delta \mathbf{m}_i \cdot \mathbf{H}}{k_B T}\right)$ (see text).

Fig. 3 : **(a)** Temperature dependence of the average life-times for a fluctuator activated around 109K. Note that by increasing the temperature the high-resistance state is stabilized and the low-resistance state become less probable. The solid lines are a guide to the eye. **(b)** Temperature dependence of the deduced activation energies for both states (we used for $\tau_{0,i}$ the values found at low temperature, see text). Insert: free energy functional $F(\sigma,T) = a(T-T_0)\sigma^2 + b\sigma^3 + c\sigma^4$, versus a configuration parameters $\sigma$ at different temperatures ($\sigma$=0 corresponding to the high-resistance state). a, b, c and $T_0$ are fitting parameters chosen to describe simultaneously the temperature dependence of the activation energies $E_{up}$ and $E_{down}$ (solid lines in panel (b)). Detailed description of the fitting procedure will be published elsewhere.

Fig 4 : Temperature dependence of the power spectral density at 10Hz normalized to the applied voltage. Inset: the resistivity versus temperature is presented. Note the 4 order of magnitude increase of the noise level between 4K and 100K, and different locations of the peaks of noise level and resistivity.

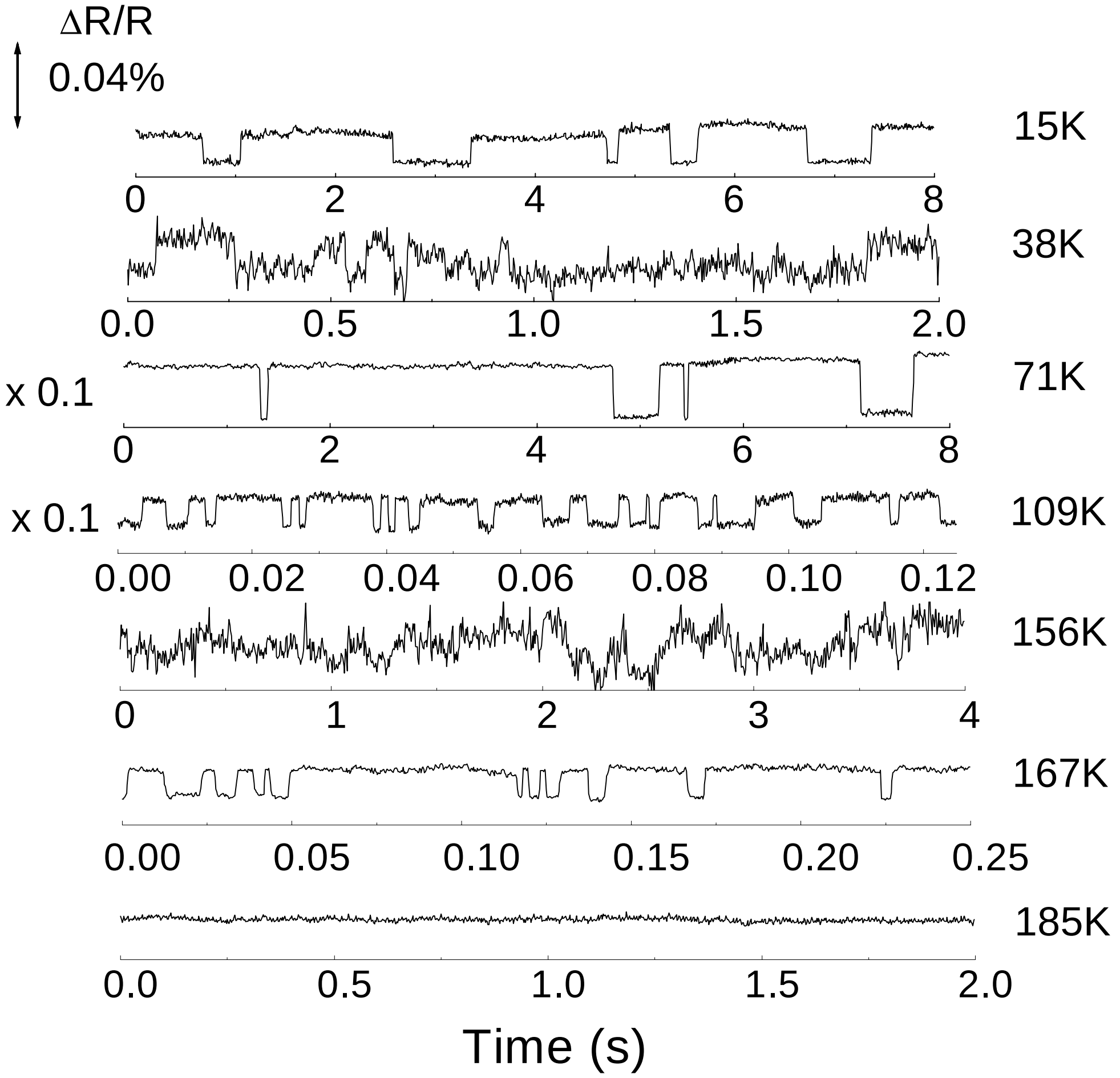

B. Raquet *et al.* : Fig. 1

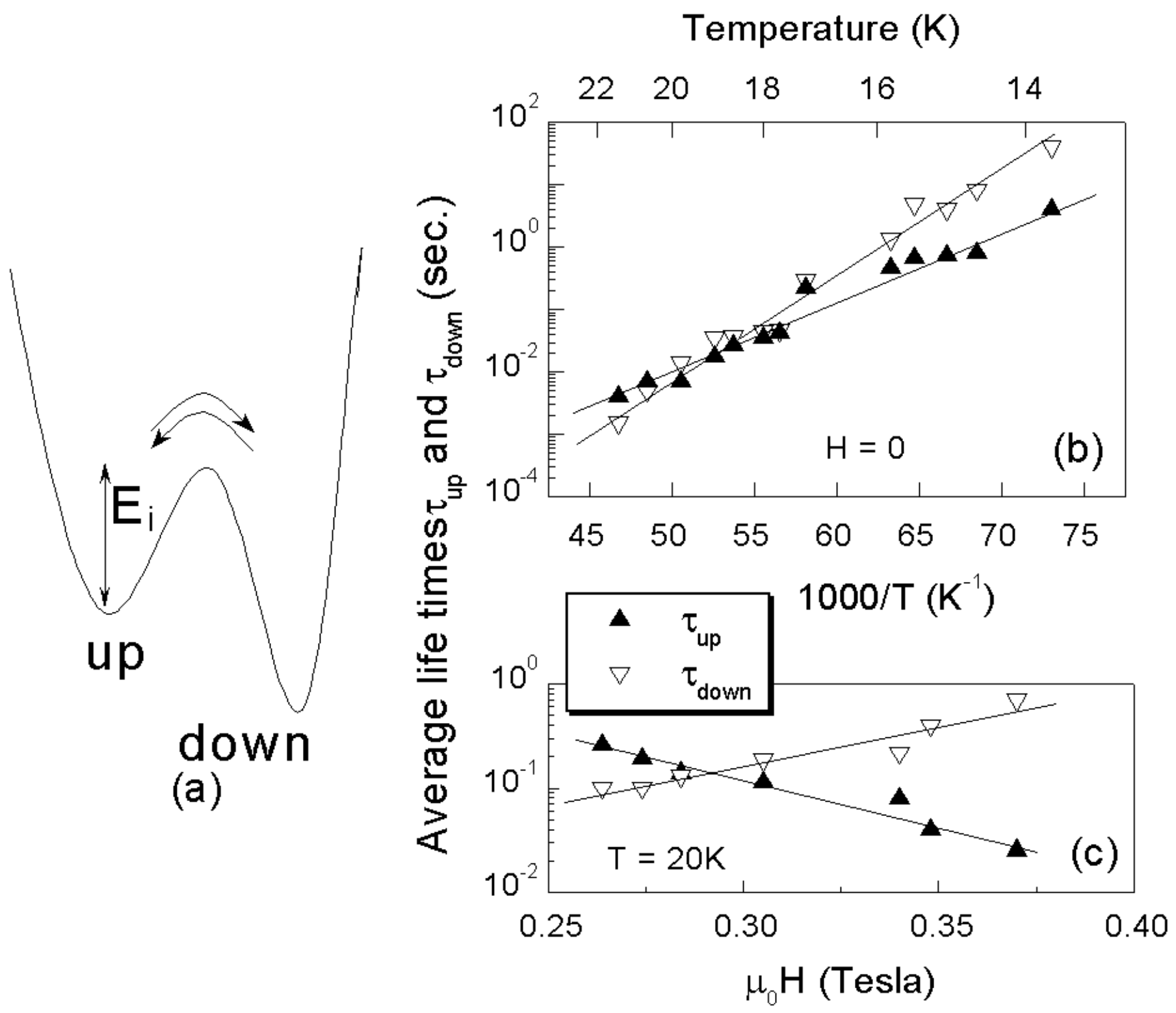

B. Raquet *et al.* : Fig. 2

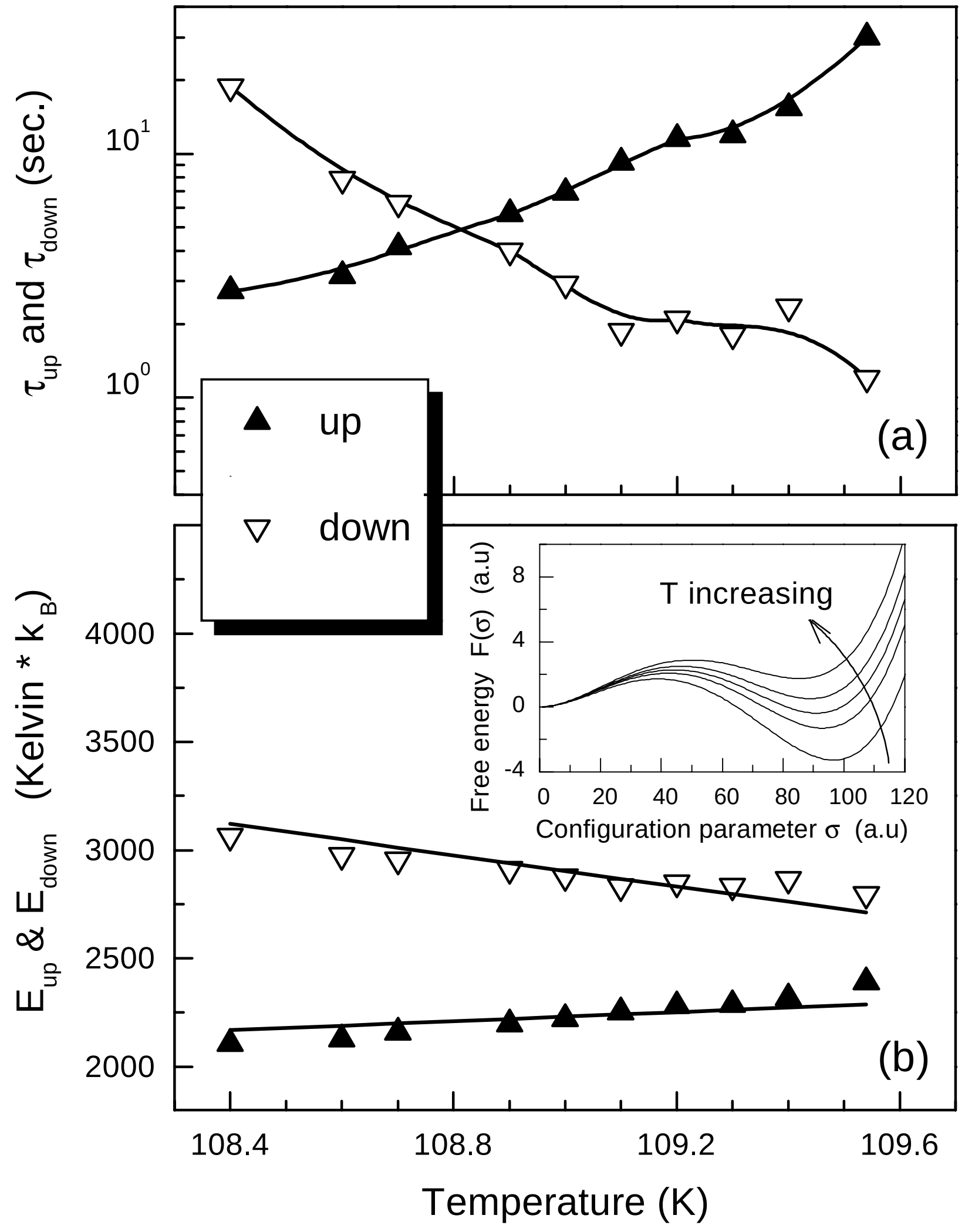

B. Raquet *et al.* : Fig. 3

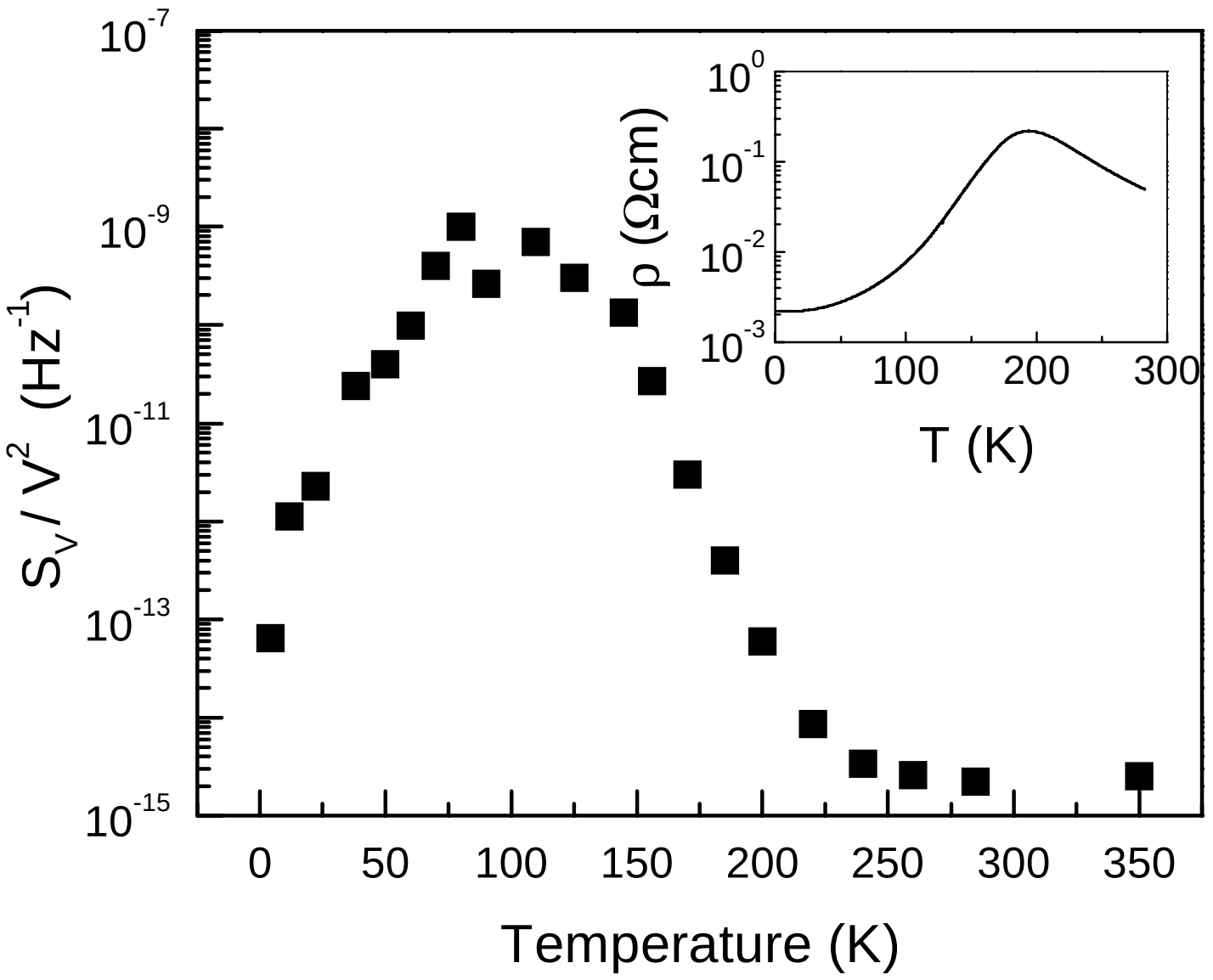

B. Raquet *et al.* : Fig. 4